\documentclass[a4paper,10pt]{article}
\usepackage[utf8x]{inputenc}
\usepackage{syntonly}
\usepackage{srcltx}
\usepackage{indentfirst}	
\usepackage{vmargin}
\usepackage{proof}

\usepackage{amsmath,amsfonts,amssymb,amsthm,mathtools}
\usepackage{latexsym}
\usepackage{graphicx}
\usepackage{color}
\usepackage{tikz}
\usetikzlibrary{arrows,shapes}
\tikzstyle{format} = [draw, thin, fill=gray!20]
\tikzstyle{elec} = [line width=1pt,draw=black!80]
 
\tikzstyle{elecred} = [line width=1pt,draw=black!80]

\usepackage{setspace}                    
\usepackage{txfonts}                     

\theoremstyle{definition}
\newtheorem{defn}{Definition}

\theoremstyle{plain}

\newtheorem{ex}{Example}
\theoremstyle{plain}

\date{}
\title{Formal Analysis of UMTS Privacy}
\author{\rm M. Arapinis, 
L. I. Mancini,
E. Ritter,
M. D. Ryan\\
University of Birmingham}

\begin{document}

\maketitle
\begin{abstract}
The ubiquitous presence of mobile communication devices and the continuous development of mobile data applications, which results in high level of mobile devices' activity and exchanged data, often transparent to the user, makes privacy preservation an important feature of mobile telephony systems.
We present a formal analysis of the UMTS Authentication and Key Agreement protocol, using the applied pi-calculus and the \texttt{ProVerif} tool. We formally verify the  model with respect to privacy properties. We show a linkability attack which makes it possible, for individuals with low-cost equipment, to trace UMTS subscribers. The attack exploits information leaked by poorly designed error messages. 
\end{abstract}
\section{Introduction}

While most mobile phone users accept that the network operator can
track their physical movements, few would be happy if an arbitrary
third party could do so. Such a possibility would enable all kinds of
undesirable behaviour, ranging from criminal stalking and harassment
to more mundane monitoring of spouse or employee movements. For this
reason, UMTS mobile phone protocols have been designed to prevent third
parties from identifying wireless messages as coming from a particular
mobile phone. The protocols include cryptography and the use of
temporary identifiers (such as the \emph{Temporary Mobile Subscriber
  Identity} (TMSI)) in an effort to achieve the aim of untrackability
by third parties.

Unfortunately, it is known that ``IMSI catchers'' can be used to force
a mobile phone to reveal its permanent \emph{International Mobile
  Subscriber Identity} (IMSI) \cite{Fox97:IC, Strobel07:IC}.  The
possibility of such a breach of IMSI confidentiality is acknowledged
in the UMTS protocol specifications \cite{TS33102}. Until fairly recently,
implementing an IMSI catcher required the specialised equipment found
only in network base stations. However, the cost of such devices is
becoming more and more affordable thanks to software emulation.

In this paper, we identify a further problem with the key
establishment protocols used in modern networks. We show that if an
adversary can capture a message to a phone that authenticates the
phone by its TMSI, then the adversary can from then on distinguish its
interactions with that phone from any other. To achieve this, the
adversary replays the captured message. Our attack exploits the fact
that the victim's phone will reply with subtly different error
messages, depending on whether the replayed request is associated with
it or with a different phone. In contrast with the IMSI catcher, our
attack is not known to the protocol designers. 

It is straightforward to modify the protocol to avoid the IMSI catcher
attack, by using public key encryption. Avoiding the error message
attack is more subtle, because the attack itself is more subtle.  We
propose a modification of the protocol.  We model the key
establishment protocol in \texttt{ProVerif}, and we demonstrate that \texttt{ProVerif}
can find the attack based on the differing error messages. Then, we
propose a simple fix, which prevents the attacker from distinguishing
between the reply to a replayed message for the victim's phone from
the reply to a replayed message from some other phone. We demonstrate
that \texttt{ProVerif} does not find the attack in the revised version, 
and actually can prove that there are no distinguishing attacks.

Unfortunately, our model is an approximation of the underlying
protocol, in a precise way. The underlying protocol uses exclusive-or
as a means to perform a symmetric-key encryption. We do not model this
encryption; indeed, we allow an adversary to see both the key and the
plaintext.  \texttt{ProVerif}'s exhibition of the attack and its absence in the
fixed protocol are both relative to this approximation. The
approximation is plainly incorrect for secrecy properties; however, the privacy
property at hand is expressed as the attacker's inability to
distinguish two situations, and intuitively, the approximation is a
correct abstraction for that property. In the approximate model, the
attacker has strictly more information with which to make the distinction.
Formally proving that the approximation is an abstraction is left as further work.

\subsection{Related Work}
\paragraph{Known Attacks on UMTS}\

Previous work analysing UMTS security exploits the
vulnerabilities which are propagated from GSM to UMTS when providing
inter-operability between the two systems. Most of the reported
attacks take advantage of well-known weaknesses of the GSM
authentication and key agreement protocol, such as {\bf lack of mutual
authentication and use of weak encryption}. These attacks allow an
active attacker to violate the user identity confidentiality, to
eavesdrop outbound communications and to masquerade as a legitimate
subscriber obtaining services which will be billed on the victim's
account. For example, Meyer and Wetzel \cite{Meyer04:MMA} present an attack which
allows the attacker to impersonate a base station and eavesdrop a
victim's mobile station communications. An even more disrupting attack
allowing impersonation of a victim's mobile station and service theft
is explained by Ahmadian et al. in \cite{Ahmadian09:NAUMTSNA}.

To the best of our knowledge, the only attack that does not rely on GSM/UMTS interoperability
has been presented by Zhang and Fang  in \cite{Zhang05:SAE3GAKA}.
The attack is a variant of the false base station attack and takes advantage of the fact that 
{\bf the mobile station does not authenticate the serving network}. It allows
the redirection of the victim's outgoing traffic to a different
network, let's say a network which uses a weaker encryption algorithm
or one which charges higher rates than the victim's one.

\paragraph{AKA Protocol Security: Formal Proofs and New Proposals }\

The UMTS authentication and key agreement protocol in its
pure form (i.e.\ with no GSM support) has been {\bf formally proved} to meet
some of the specified security requirements \cite{TR33902}, such as authentication and confidentiality.
However, the framework used for the formal analysis does not capture Zhang and Fang's attack.
Moreover, unlinkability and anonymity properties, which are the focus of our work, have not been analysed.

A new framework for authentication has been proposed to take into
account subscribers' privacy with respect to both the serving and the
home network \cite{Koein06:LPCSAS}.  Other recently published papers
also propose new authentication protocols to enhance UMTS security and
privacy \cite{Koein06:LPCSAS,Zhang05:SAE3GAKA}.
However, the adoption of the enhanced authentication protocols would require considerable changes to the current security architecture.
Unlike our work, these works, do not rely on a formal study of the AKA protocol or present a formal verification of the proposed protocols.

\section{Introduction to UMTS}
Universal Mobile Telecommunications System (UMTS) is a mobile telephony standard specified and maintained by the Third Generation Partnership Project (3GPP). UMTS was introduced in 1999 to offer a better support for mobile data applications increasing the data rate and lowering the costs of mobile data communications. Furthermore, UMTS offers an improved security architecture with respect to previous mobile communication systems such as GSM (Global System for Mobile Communication). In the following sections, we will introduce the UMTS network architecture and we will describe in more detail its security features and the Authentication and Key Agreement (AKA) protocol, which is the main building block of UMTS security. We will then present a linkability attack which enables tracing UMTS subscribers thanks to the exploitation of the AKA protocol error messages.
\subsection{UMTS Architecture}
\begin{figure}[tp]
\begin{center}
	\includegraphics[width=0.70\textwidth]{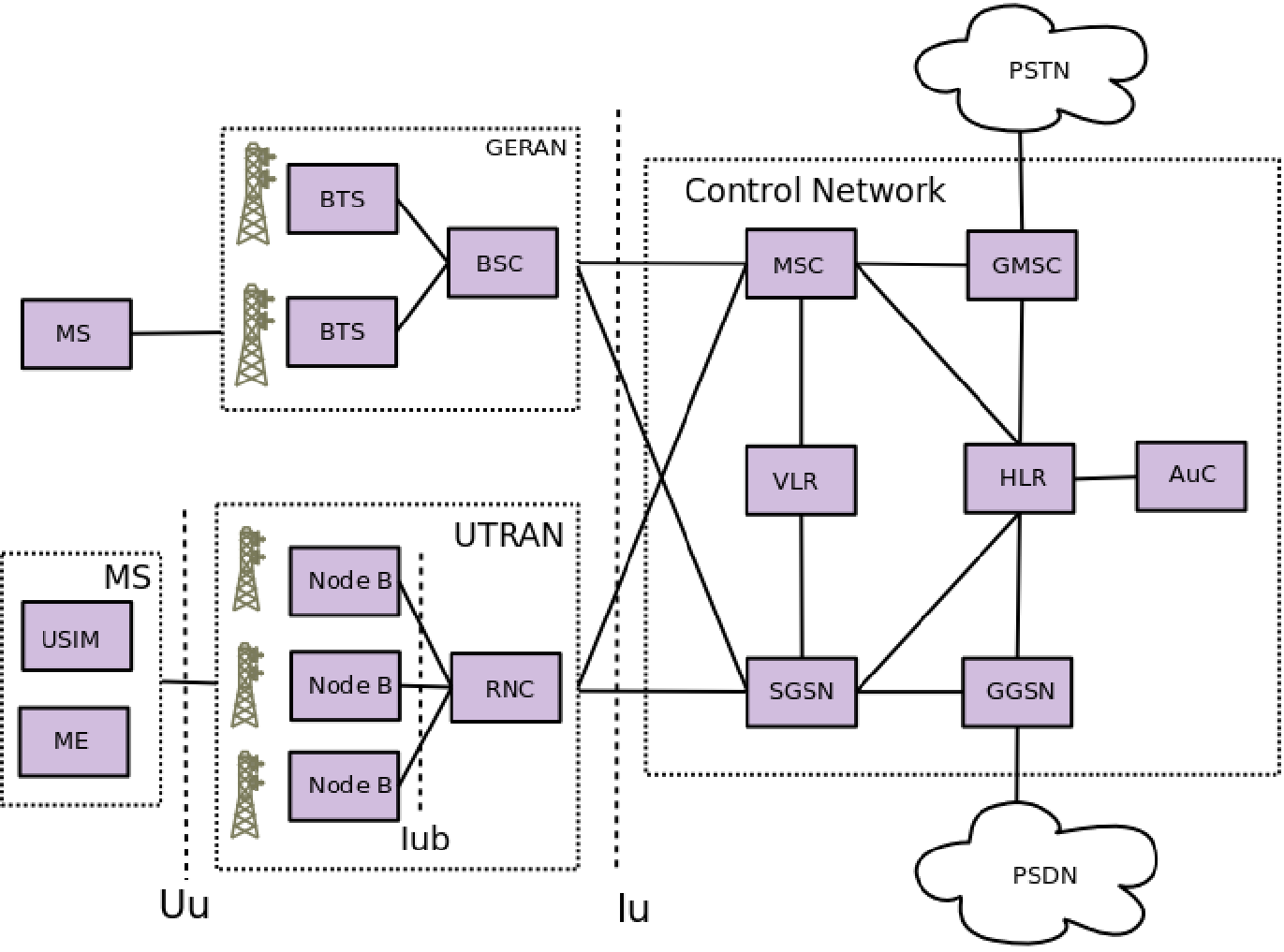} 
\end{center}
\caption{UMTS Architecture}
\label{UMTSArch}
\end{figure}

The network architecture of the UMTS system, depicted in figure \ref{UMTSArch}, is built on the pre-existing  voice and data mobile networks, by namely GSM and GPRS (General Packet Radio Service) networks, respectively.  The user side of the network consists of Mobile Stations (MS). Each mobile station comprises a Mobile Equipment (ME) such as a mobile phone, together with a Universal Subscriber Identity Module (USIM), which identifies the user as a legitimate subscriber within a mobile telephony operator network. To access the services offered by a mobile operator, a MS connects through radio communication technology to the UTRAN (UMTS Terrestrial Radio Access Network) or GERAN (GSM/EDGE Radio Access Network). The latter one is a GSM/GPRS access network. 

A mobile station directly communicates with a Base Transceiver Station or Node B which covers the area the MS is located in. One or more Node B are connected with a Radio Network Controller (RNC), defining a cell. A group of cells forms a Location Area. RNCs manage the radio resources and inter-cell handover and are the interface between the mobile station and the core network. The core network offers circuit-switch and packet-switch services. The Mobile Switching Centre (MSC) and Gateway Mobile Switching Centre (GMSC) offer, respectively, inter and intra-network circuit-switching domain services and interface the UMTS with the traditional fixed telephony network. The Serving GPRS Support Node (SGSN) and the Gateway Serving GPRS Support Node (GGSN) offer, respectively, inter and intra-network packet-switching domain services as well as connecting the UMTS network with the internet. 

Within the core network, the Home Location Register (HLR) stores permanent sensitive information of UMTS subscribers such as identity, service profile, and activity status. These informations are linked to the USIM and recorded when stipulating a contract with the mobile operator. UMTS offers roaming capabilities between different network operators, meaning that a mobile station can be connected to a visited network, called the Serving Network (SN), which might be different from the subscriber's Home Network (HN). 

Each subscriber has a long term identifier IMSI (International Mobile Subscriber Identity) that is stored in the USIM, and a temporary identifier TMSI (Temporary Mobile Subscriber Identity), allocated by the serving network. The purpose of the TMSI is to protect the subscriber identity privacy. To limit the use of the IMSI a subscriber identifies himself using the TMSI whenever it is possible. To avoid user traceability, the TMSI is periodically changed through the TMSI reallocation procedure. 

The Visitor Location Register (VLR) stores temporary information about subscribers visiting a given location area of the serving network and maintains TMSI/IMSI associations. The network operator and the subscriber share a long term secret key used for authentication purposes. This key is stored in the USIM. The Authentication Centre (AuC) is a protected database storing association between subscriber identities (IMSI) and long-term keys.

\subsection{UMTS Security}
We will consider a simplified network architecture so to be able to abstract from the network complexity and concentrate on the communication protocols taking place between a Mobile Station (MS) and the Serving Network (SN). 

When referring to a Serving Network we mean both the UTRAN/GERAN Base Station that the MS is directly communicating with, and the complex structure of databases and servers connected with it and forming the UMTS core network.  

When a mobile station connects to a network, a temporary identity, TMSI (Temporary Mobile Subscriber Identity) is assigned to it and is used, instead of the IMSI, to identify the MS. To avoid mobile station traceability by third parties, the UMTS standard requires periodic updates of the temporary identity to be performed.

UMTS communication system aims to provide authentication, confidentiality, and user identity confidentiality \cite{TS33102}. In particular, the UMTS standard defines the following two properties concerning the subscribers privacy:
\begin{itemize}
	\item User identity confidentiality: the property that the permanent user identity (IMSI) of a user to whom a services is delivered cannot be eavesdropped on the radio access link;
	\item User untraceability: the property that an intruder cannot deduce whether different services are delivered to the same user by eavesdropping on the radio access link.
\end{itemize}

To achieve the above mentioned security properties, UMTS relies mainly on the Authentication and Key Agreement (AKA) protocol, on ciphering and integrity checking of the confidential data transmitted on the radio channel, and on the use of temporary identities. The AKA protocol allows MS and SN to establish a pair of shared session keys: a ciphering key and an integrity key, to be used to ensure the secrecy and integrity of the subsequent communications.

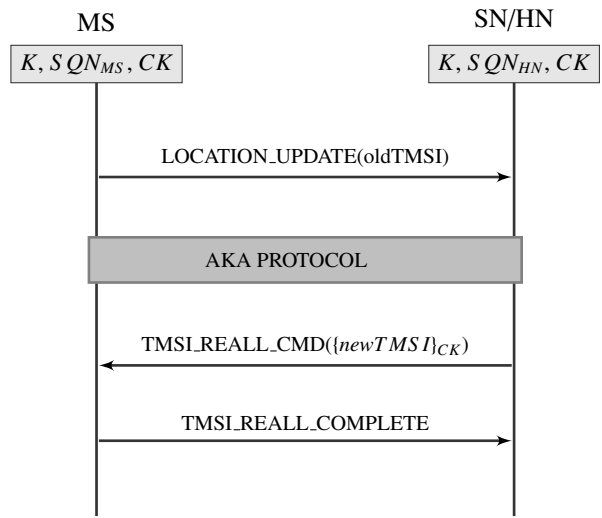
\begin{figure}[tp]
\begin{tikzpicture}[node distance=4cm, auto,>=latex', thick]
	
    \path[use as bounding box] (0.8,3) rectangle (16,-3.5);
    \draw[elec] (5.5,2.5) node[format](ms){\small $K$, $SQN_{MS}$, $CK$}-- (5.5,-3.5);
    \draw[elec] (11,2.5) node[format](sn){\small $K$, $SQN_{HN}$, $CK$}-- (11,-3.5);
	\node [above=3mm] at (sn){SN/HN};
    \node [above=3mm] at (ms){MS};
	\node at (11.1,1)(startreq){};
    \node at (5.4,1)(endreq){};
    \path[elec,<-](startreq) edge node[above] {\footnotesize LOCATION\_UPDATE(oldTMSI)} (endreq);

    \draw[elec,color=gray,fill=lightgray] (5.4,0.2) rectangle(11.1,-0.4);
	\node at (8.0,-0.1)(auth){\footnotesize AKA PROTOCOL};

    \node at (5.4,-1.5)(startreq){};
    \node at (11.1,-1.5)(endreq){};
    \path[elec,<-](startreq) edge node[above] {\footnotesize TMSI\_REALL\_CMD$(\{newTMSI\}_{CK})$} (endreq);

    \node at (5.4,-2.5)(startresp){};
    \node at (11.1,-2.5)(endresp){};
    \path[elec,->](startresp) edge node[above] {\footnotesize TMSI\_REALL\_COMPLETE} (endresp);
  
\end{tikzpicture}
\caption{TMSI Reallocation procedure}
\label{tmsi}
\end{figure}

As previously mentioned, mobile stations are identified using temporary identities. New temporary identities are assigned by the network using the TMSI reallocation procedure, which consists of a message containing the new TMSI, followed by an acknowledgement message. To avoid the linkability of consecutively used temporary identities, the new TMSI is encrypted using the session key $CK$ established during the execution of the AKA protocol. The standard requires a new TMSI to be allocated at least at each change of location area \cite{TS24008}. The TMSI reallocation is usually performed in conjunction with other procedures as for example location updating. This scenario is depicted in figure \ref{tmsi}.
\subsection{The UMTS Authentication and Key Agreement Protocol}
The Authentication and Key Agreement (AKA) protocol aims to achieve the mutual authentication of MS and SN, and to establish shared session keys to be used to secure the subsequent communications. The session keys are not exchanged during the protocol but computed locally by the MS and the SN. According to \cite{TS33102}, the authentication procedure is always initiated by the SN for the purpose of:
\begin{itemize}
 \item Checking whether the identity provided by the MS is acceptable or not.
 \item Providing parameters enabling the MS to calculate a new UMTS ciphering key.
 \item Providing parameters enabling the MS to calculate a new UMTS integrity key.
 \item Allowing the MS to authenticate the network.
\end{itemize}
The MS and the Home Network (HN) share a secret cryptographic key $K$, assigned to the subscriber by the network operator and loaded on the USIM card. 
The secret key allows MS and HN to compute, off-line, shared ciphering and integrity keys to be used for encryption and integrity check of the communication. 

\begin{figure}
\begin{tikzpicture}[node distance=4cm, auto,>=latex', thick]

    \path[use as bounding box] (0.8,3) rectangle (16,-7);
    \draw[elec] (5.5,2.5) node[format](ms){ $K$, $SQN_{MS}$}-- (5.5,-7);
    \draw[elec] (11,2.5) node[format](sn){ $K$, $SQN_{HN}$}-- (11,-7);
	\node [above=3mm] at (sn){SN/HN};
    \node [above=3mm] at (ms){MS};
	\node at (11.1,-1)(startreq){};
    \node at (5.4,-1)(endreq){};
    \path[elec,->](startreq) edge node[above] {\footnotesize AUTH\_REQ$(RAND,AUTN)$} (endreq);
    \draw[elec] (11,0.8) node [format, fill=white](sn_pre){
			\parbox{3.6cm}{
				{\footnotesize
					new $RAND$\\
					$AK\leftarrow f5_K(RAND)$\\
					$MAC\leftarrow f1_K(SQN_{HN}||RAND)$\\
					$AUTN\leftarrow SQN_{HN}\oplus AK||MAC$
				}
			}
	 };
	 \draw[elec] (5.5,-3) node [format, fill=white](sn_pre){
			{\parbox{3.6cm}
				{\footnotesize
				$AK\leftarrow f5_K(RAND)$\\
				$XMSG||XMAC\leftarrow AUTN$\\
				$SQN_{HN}\leftarrow XMSG \oplus AK$\\
				$MAC\leftarrow f1_K(SQN_{HN}||RAND)$\\
				Check: $MAC=XMAC$\\
 			   Check: $SQN_{HN}>SQN_{MS}$\\
				$RES\leftarrow f2_K(RAND)$
				}
			}
	 };
    \node at (5.4,-5)(startresp){};
    \node at (11.1,-5)(endresp){};
    \path[elec,->]  (startresp) edge node[above] {\footnotesize AUTH\_RES$(RES)$} (endresp);
	 \draw[elec] (5.5,-6) node [format, fill=white](sn_pre){
			{\parbox{3.6cm}
				{\footnotesize
				$CK\leftarrow f3_K(RAND)$\\
				$IK\leftarrow f4_K(RAND)$
				}
			}
	 };
    \draw[elec] (11,-6) node [format, fill=white](sn_pre){
		\parbox{3.6cm}
			{\footnotesize
  			Check: $RES=f2_K(RAND)$\\
			$CK\leftarrow f3_K(RAND)$\\
			$IK\leftarrow f4_K(RAND)$
			}
	};
\end{tikzpicture}
  \caption{Authentication and Key Agreement protocol}
  \label{AKAP}
\end{figure}
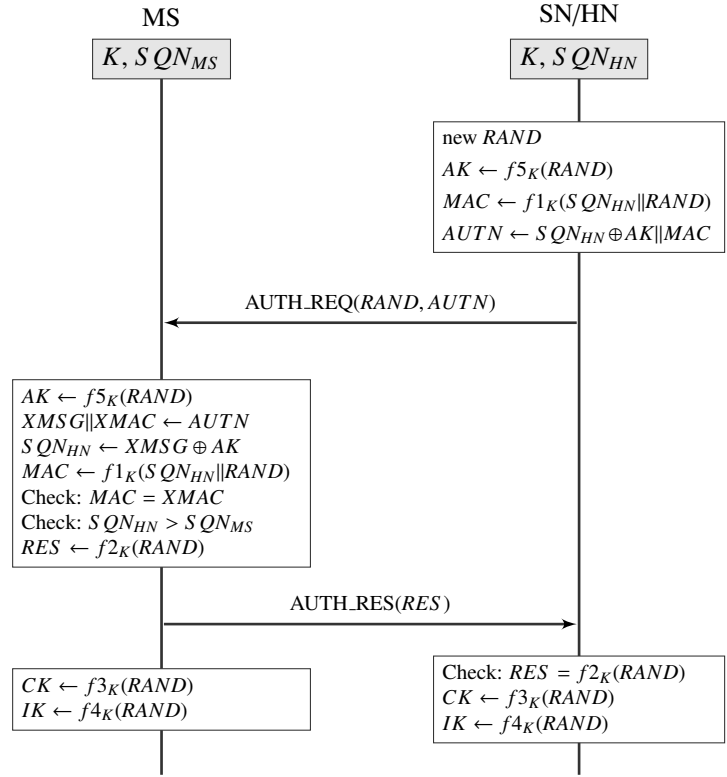
The AKA protocol consists in the exchange of two messages: the authentication request and the authentication response. In figure \ref{AKAP}, a successful execution of the AKA protocol is shown. The SN initiates the authentication and key agreement procedure as follows:
\begin{itemize}
	\item Before sending an authentication request to the MS, the serving network requests authentication data from the mobile station's home network (HN). The home network returns an authentication vector containing a random challenge $RAND$, the authentication token $AUTN$ computed from it, the expected authentication response $f2_K(RAND)$, the integrity key $IK$ and the encryption key $CK$. The  functions, $f1, f2, f3, f4,$ and $f5$, used to compute the authentication parameters are cryptographic functions computed using the shared key $K$ and are defined in the UMTS standard. The authentication function $f1$ is used to calculate the MAC (Message Authentication Code), $f2$ is used to produce the authentication response parameter $RES$; the key generating functions, $f3, f4,$ and $f5$ are used to generate the ciphering, $CK$, the integrity, $IK$, and the anonymity key, $AK$, respectively.
	\item The SN sends the authentication challenge RAND and the authentication token $AUTN$ to the MS. The authentication token contains a MAC which is computed applying the function $f1$ to the concatenation of the random number with a sequence number, $SQN_{HN}$, generated by the HN/AuC using an individual counter for each user. A new sequence number is generated either by increment of the counter or through time based algorithms as defined in \cite{TS33102}. The sequence number $SQN_{HN}$ identifies the authentication vector and allows the MS to verify the freshness of the authentication request (see figure \ref{AKAP}).
	\item When the MS receives the $(RAND, AUTN)$ message it first retrieves the sequence number ($SQN_{HN}$), verifies the MAC ($MAC = XMAC$). This step ensure the MAC was generated by the network using the shared key $K$. The mobile station stores the greatest sequence number used for authentication, $SQN_{MS}$. This value is used to check the freshness of the authentication request ($SQN_{MS}<SQN_{HN}$). 
	\item The MS computes the ciphering key, the integrity key and the authentication response $RES$ message and sends $RES$ to the network. 
	\item The SN authenticates the MS verifying whether the received response is equal to the expected one ($RES=f2_K(RAND)$). 
\end{itemize}

After successful authentication, the SN sends a security mode command message to the MS, indicating which one of the allowed algorithms to use for ciphering and integrity checking of the subsequent communications.

The authentication procedure can fail on the MS side either because the MAC verification failed, or because the received sequence number is not in the correct range with respect to the sequence number stored in the MS ($SQN_{MS}$). In the former case, the MS sends an authentication failure message indicating \textit{Mac Failure} as the failure cause. In the latter case, the MS sends an authentication failure message containing the $AUTS$ parameter and indicating \textit{Synchronization Failure} as the failure cause. The AUTS parameter contains the sequence number $SQN_{MS}$, to allow the SN to perform a re-synchronization and a MAC for authenticity and integrity purposes.

The authentication procedure can fail on the SN side because the $RES$ verification failed, in this case the SN sends an authentication reject message.

\begin{figure}
\begin{tikzpicture}[node distance=4cm, auto,>=latex', thick]
	
    \path[use as bounding box] (0.8,3) rectangle (16,-5.5);
    \draw[elec] (5.5,2.5) node[format](ms){\small $K$,$SQN_{MS}$}-- (5.5,-5.5);
    \draw[elec] (11,2.5) node[format](sn){\small $K$,$SQN_{HN}$}-- (11,-5.5);
	\node [above=3mm] at (sn){SN/HN};
    \node [above=3mm] at (ms){MS};
	\node at (11.1,1)(startreq){};
    \node at (5.4,1)(endreq){};
    \path[elec,->](startreq) edge node[above] {\footnotesize AUTH\_REQ$(RAND,AUTN)$} (endreq);
    \node at (5.4,-2.7)(startresp){};
    \node at (11.1,-2.7)(endresp){};
    \path[elecred,->](startresp) edge node[above] {\footnotesize AUTH\_RESPONSE$(RES)$} (endresp);
	 \draw[elec] (5.5,-0.8) node [format, fill=white](sn_pre){
			\parbox{3.7cm}{
				{\footnotesize
 				if $MAC\neq XMAC$ then \\
				$RES\leftarrow MAC\_FAIL$\\
				else if $SQN_{HN}<=SQN_{MS}$ then\\
				$RES\leftarrow SYNCH\_FAIL(AUTS)$\\
				else $RES\leftarrow f2_K(RAND)$
				}
			}
	 };
	 \draw[elec] (11,-4.3) node [format, fill=white](sn_pre){
			\parbox{3.7cm}{
				{\footnotesize
  				if $RES$=SYNCH\_FAIL then\\
				Resynch\\
				else if $RES\neq f2_K(RAND)$ then\\
				Reject Authentication
				}
			}
	 };
\end{tikzpicture}
\caption{AKA Failure Messages}
\label{AKAfail}
\end{figure}
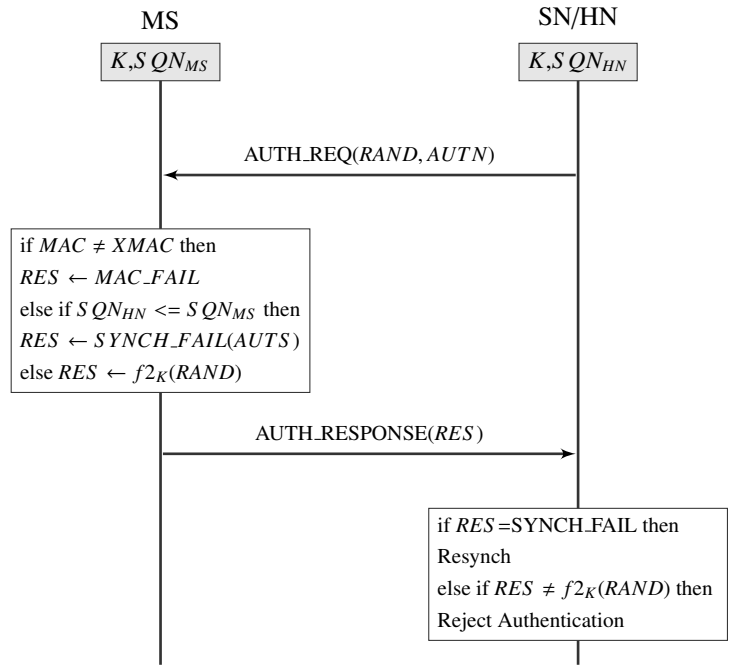

For simplicity, we are not considering here failure situations in which the network initiates other procedures to recover from an authentication failure. Figure \ref{AKAfail} depicts the error messages possibly occurring during the authentication procedure.

\section{Linkability Attack}
\label{linkatt}
\begin{figure}
\begin{tikzpicture}[node distance=4cm, auto, thick]

    \path[use as bounding box] (2.5,3) rectangle (16,-2.3);

    \draw[elec] (5,2.5) node[format](ms){\small $K$,$SQN_{MS_v}$}-- (5,-2.3);
    \draw[elec] (10,2.5) node[format](mim){\small $RAND, AUTN$}-- (10,-2.3);
    \draw[elec] (15,2.5) node[format](sn){\small $K$,$SQN_{HN}$}-- (15,-2.3);
	\node [above=3mm] at (sn){SN/HN};
	\node [above=3mm] at (mim){Attacker};
    \node [above=3mm] at (ms){MS};
	
	\node at (15.1,1.5)(startreq){};
    \node at (4.9,1.5)(endreq){};
    \path[elec,->](startreq) edge node[above] {\footnotesize AUTH\_REQ$(RAND,AUTN)$} (endreq);
 	\node at (4.9,0.7)(startresp){};
    \node at (15.1,0.7)(endresp){};
    \path[elec,->](startresp) edge node[above] {\footnotesize AUTH\_RES$(RES)$} (endresp);
	 \draw[elec] (5,-0.1) node[format](ms){\small $K$,$SQN'_{MS_v}$}-- (5,-2.3);
	\node at (10.1,-0.9)(startreq1){};
    \node at (4.9,-0.9)(endreq1){};
    \path[elec,->](startreq1) edge node[above] {\footnotesize AUTH\_REQ$(RAND,AUTN)$} (endreq1);
	\node at (4.9,-1.7)(startresp1){};
    \node at (10.1,-1.7)(endresp1){};
    \path[elecred,->](startresp1) edge node[above] {\footnotesize AUTH\_RES$(SYNCH\_FAIL)$} (endresp1);	
	\node [right] at (10,-1.7) {\footnotesize\parbox{4cm}{\flushleft
 			This is $MS_v$!}};
   
\end{tikzpicture}
\caption{Linkability attack}
\label{attack}
\end{figure}
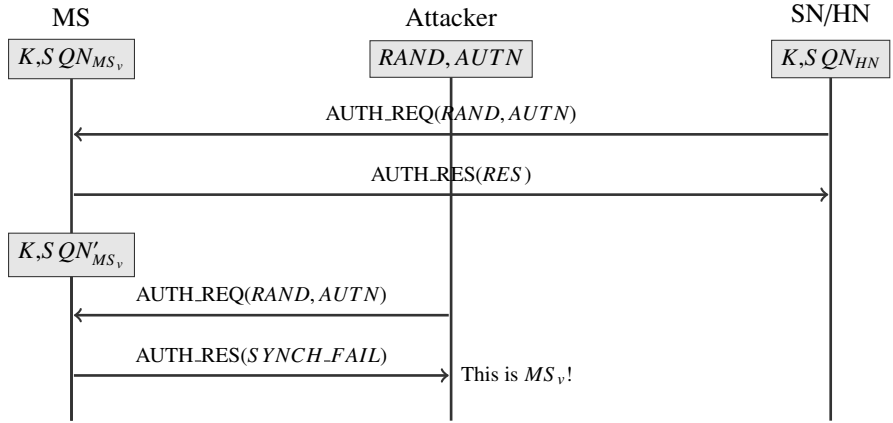

Despite the use of temporary identities to avoid the linkability of UMTS subscribers, the execution of the AKA protocol enables an active attacker to trace UMTS subscribers. An active attacker could intercept an authentication request message containing the pair $(RAND, AUTN)$ sent by the SN to a victim mobile station $MS_v$. The captured authentication request can be replayed by the adversary to check the presence of $MS_v$. In fact, thanks to the error messages, the adversary can distinguish any mobile station from the one the authentication request was originally sent to. On reception of the replayed $(RAND, AUTN)$, the victim mobile station, $MS_v$, successfully verifies the MAC and sends a synchronization failure message. While, the MAC verification fails when executed by any other mobile station and as a result a MAC failure message is sent. The implementation of few false base stations would then allow an attacker to trace the movements of a victim mobile station, resulting in a breach of the subscriber untraceability. The proposed attack is shown in figure \ref{attack}. 

Intuitively, the UMTS AKA protocol linkability problem can easily be solved by simply sending the same error message for any kind of failure occurring.
In the following sections, we introduce the theoretical framework we used to automatically verify the unlinkability and anonymity of this simple fix. In addition, we illustrate the UMTS AKA protocol model and discuss the verification results.

\section{Formal Analysis of Privacy-related Properties}
In the following sections we present the \texttt{ProVerif} process language. We introduce the model of UMTS AKA protocol and the formalization of the privacy properties under analysis. Further, we discuss the result of the verification with the \texttt{ProVerif} tool of both the original protocol and the fixed version.
\subsection{ProVerif Process Language}
The \texttt{ProVerif} process language is similar to the applied pi-calculus, which is a formal language, introduced by Abadi and Fournet \cite{Abadi01:MVNNSC}, for modelling concurrent processes and ease the modelling of cryptographic protocols. The \texttt{ProVerif} process language makes it possible to automatically verify protocol models written in the language, using the \texttt{ProVerif} tool \cite{Blanchet05:AVSESP}.

Cryptographic primitives are modelled as functions and messages are represented by \textit{terms} of the \texttt{ProVerif} process language built over an infinite set of names, an infinite set of variables and the set of considered cryptographic primitives. A function symbol with arity 0 is a constant symbol. Functions are distinguished in two categories: constructors and destructors. We write $f$ for constructors and $g$ for destructors. Terms are variables, names and application of constructors and are defined by the following grammar: 
\begin{displaymath}
\begin{array}{l l l}
  L,M,N,T &::=  & \textnormal{Terms}  \\
     & a,b,c,\dots   &  \textnormal{Name}\\
  &x,y,z,\dots   & \textnormal{Variable} \\
  &f(M_1,\dots,M_l)& \textnormal{Function application}\\
   &\multicolumn{2}{c}{\textnormal{where } f\in\Sigma \textnormal{ and }l\textnormal{ matches the arity of }f}\\
\end{array}
\end{displaymath}
Destructors are defined by a set of reduction rules of the form $g(M_1,\dots,M_l)\rightarrow M$ which can be applied by processes to manipulate terms.
\begin{ex}
Using functions and reduction rules we can define cryptographic functions, for example, let $\Sigma=\{\texttt{senc}/2, \texttt{pub}/1, \texttt{aenc}/3, \texttt{xor}/2, \texttt{0}/0,\}$, and consider the reductions
\begin{displaymath}
\begin{array}{l l l}
\texttt{sdec}(k, \texttt{senc}(k, m))&\rightarrow& m\\
 & & \\
\texttt{adec}(k, \texttt{aenc}(\texttt{pub}(k),m,r))&\rightarrow& m\\
 & & \\
\texttt{xor}(\texttt{xor}(x, y), z)&\rightarrow&\texttt{xor}(x, (\texttt{xor}(y, z))\\
\texttt{xor}(x, y)&\rightarrow&\texttt{xor}(y, x)\\
\texttt{xor}(x, \texttt{0})&\rightarrow& x\\
\texttt{xor}(x, x)&\rightarrow& \texttt{0}\\
\end{array}
\end{displaymath}
In this example, \texttt{senc} and \texttt{aenc} model, respectively, symmetric and non deterministic asymmetric encryption. The first reduction rule allows to decrypt an encrypted message, $m$, given the knowledge of the encryption key $k$, hence it defines symmetric encryption. While, the second rules allows to decrypt a message, $m$, encrypted using a public key, $pub(k)$, and a random value, $r$, given the knowledge of the corresponding private key $k$. Therefore, it defines non deterministic public key encryption.

The function \texttt{xor} is described by a set of reductions modelling its algebraic properties and a constant \texttt{0} representing the value 0.
\end{ex}
The grammar for \textit{processes} of the \texttt{ProVerif} process language is the following:
\begin{displaymath}
\begin{array}{l l l}
  P,Q,R ::=  & \textnormal{plain processes}  \\
  0    &\textnormal{null process}\\
  P\mid Q &\textnormal{parallel composition} \\
  !P& \textnormal{replication}\\
  \texttt{new } n;P &\textnormal{name restriction}\\
  \texttt{if } M=N \texttt{ then } P \texttt{ else } Q &\textnormal{conditional}\\
  \texttt{let } x=g(M_1,\dots ,M_n)  \texttt{ in } P &\textnormal{destructor application}\\
  \hspace{3cm}\texttt{ else } Q &\\
  \texttt{in}(M, x);P&\textnormal{message input}\\
  \texttt{out}(M,N);P& \textnormal{message output}\\
\end{array}
\end{displaymath}
The null process does nothing. The parallel composition of $P$ and $Q$ represents the parallel execution of $P$ and $Q$. The replication of a process $P$ acts like the parallel execution of unboundedly many copies of $P$. The name restriction $\nu n.P$ creates a new name $n$ whose scope is restricted to the process $P$ and then runs $P$. Note that we check for equality modulo the equational theory rather than syntactic equality of terms. 
The message input $\texttt{in}(M,x);P$ represents a process ready to input from the channel $M$, the actual message received will be substituted to $x$ in $P$, the syntactic substitution of a term $T$ for the variable $x$ in the process $P$ is denoted by $P\{^T/_x\}$. The message output $\texttt{out}(M,N);P$ describes a process ready to send a term $N$ on the channel $M$ and then to run $P$. The let construct defines a process that evaluates $g(M_1,\dots ,M_n)$ and behaves as $P$, where $x$ is substituted with the evaluation of $g(M_1,\dots ,M_n)$ in $P$, if the evaluation of $g(M_1,\dots ,M_n)$ is successful, and behaves as $Q$ otherwise. The conditional checks the equality of two terms $M$ and $N$ and then behaves as $P$ or $Q$ accordingly. 

The conditional is actually defined in terms of \texttt{let}. A destructor \texttt{equals} is defined, with the reduction $\texttt{equals}(x,x)\rightarrow x$. The \texttt{if} construct stands for $\texttt{let}\; x=\texttt{equals}(M,N)\, \texttt{in}\; P\; \texttt{else}\; Q$. We also use some syntactic sugar to allow the let construct to introduce abbreviations such as $\texttt{let }x=M\texttt{ in }P$. Moreover, we assume implicit definitions of constructors and destructors of tuples. In particular, we allow implicit application of tuple destructors using pattern matching, as for example $\texttt{let } (x_1,...,x_n) = M\texttt{ in }P \texttt{ else } Q$, that if $M$ is an $n$-tuple, executes $P$ where $x_1,\dots,x_n$ are substituted with $fst(M), snd(M),\dots, n-th(M)$, and executes $Q$ otherwise.

The definition of a process in the ProVerif language consists of a set of declarations which allow to define:
\begin{description}
 \item{\verb|free a.|} free names;
 \item{\verb|fun f/arity.|} functions and their arity;
 \item{\verb|reduc f(x,y,..)=f(w,z,..).|} the functions' reduction rules;
 \item{\verb|let A = P.|} a subprocess A, where P is defined by the grammar given above;
 \item{\verb|Process P|} the main process, where P is defined by the grammar given above. 
\end{description}
\begin{ex}\label{exproc}
We can model a system $S$ where mobile stations, $MS$, with identity, $imsi$, run along with a serving network, $SN$, with which they share a private key $k$ as the process:
\begin{verbatim}
let S = !new k; new imsi; !new sqn;(SN | MS)
\end{verbatim}
The serving network process SN executing the AKA protocol can be modelled by the following \texttt{ProVerif} process:
\begin{verbatim}
1 let SN = 
2    new rand;
3    let mac = f1(k, (rand, sqn)) in
4    let ak = f5(k, rand) in
5    let autn = (xor(sqn, ak), mac) in
6        out(c, (rand, autn));
7        in(c, xres).
\end{verbatim}
The \texttt{SN} creates a fresh random number and calculates the parameters needed during the authentication process. It sends the random and the authentication token on a public channel \texttt{c}. Finally, the SN waits for the authentication response.

A mobile station can be represented by the following process:
\begin{verbatim}
1 let MS =   
2       in(c, x);
3       let (xrand, xautn) = x in
4       let (y, xmac) = xautn in
5       let ak = f5(k, xrand) in
6       let xsqn = xor(y, ak) in
7       let mac = f1(k, (xrand, xsqn)) in
8          if xmac = mac then
9              if xsqn = sqn then
10                let res = f2(k, xrand) in
11                   out(c, res)
12             else 
13                out(c, synchFail)
14         else
15             out(c, macFail).
\end{verbatim}
The \texttt{MS} waits for an authentication request, from which it retrieves the mac and the sequence number (respectively \texttt{xmac} and \texttt{xsqn}), checks if they are equal to the locally computed ones (\texttt{mac, sqn}) and sends either the authentication response \texttt{res} or a failure message, indicating the failure cause. 
\end{ex}
\subsection{Formal Semantics}
The evaluation of terms ($\Downarrow$) of the \texttt{ProVerif} process language is defined by the following rules:
\begin{displaymath}
\begin{array}{r c l l} 
M&\Downarrow&M&\\
g(D_1,\dots,D_n)&\Downarrow& \sigma N&\\
\multicolumn{4}{l}{\hspace{5mm}\textnormal{if}\; g(N_1,\dots,N_n)\rightarrow N\;\in \Sigma \textnormal{ and }}\\
\multicolumn{4}{l}{\hspace{5mm}\sigma \textnormal{ is such that }\forall i,\; D_i\Downarrow M_i\textnormal{ and }M_i=\sigma N_i}\\
\end{array}
\end{displaymath}
The structural equivalence relation ($\equiv$) defines when syntactically different processes represent the same process, allowing to rule out uninteresting syntactic differences between processes:
\begin{displaymath}
\begin{array}{r c l l} 
P&\equiv&P\mid 0&\\
P\mid Q&\equiv&Q \mid P&\\
P\mid(Q \mid R)&\equiv&(P\mid Q)\mid R&\\
\texttt{new}\; a;\texttt{new}\; b;P &\equiv& \texttt{new}\; b;\texttt{new}\; a;P&\\
P \mid \texttt{new}\; a;Q &\equiv& \texttt{new}\; a;(P \mid Q)&\textnormal{if }a \notin fn(P)\\
P&\equiv& P&\\
\end{array}
\end{displaymath}
\begin{displaymath}
\begin{array}{l l l} 
\infer[]{Q\equiv P}{P\equiv Q} & \infer[]{P\equiv R}{P\equiv Q, Q\equiv R}&\infer[]{\texttt{new}\; a;P\equiv \texttt{new}\; a;Q}{P \equiv Q}\\
\end{array}
\end{displaymath}
The reduction relation ($\rightarrow$) describes how processes evolve. According to the first reduction rule, input and output actions on a channel $N$ can synchronize, resulting in the message $M$ to be communicated, i.e. substituted for $x$ in $Q$. The let construct tries to evaluate the term $D$, if the evaluation is successful, i.e. $D\Downarrow M$ then $M$ is substituted for $x$ in $P$ otherwise the process $Q$ is executed. The replication of $P$ can evolve to the parallel execution of an unfolded copy of $P$ running in parallel with its replication. Moreover, we have that if a process $P$ can evolve to $Q$, when in isolation, then it can perform the same step when running in parallel with another process $R$. The reduction relation is closed by name restriction and structural equivalence.
\begin{displaymath}
\begin{array}{r c l l} 
\texttt{out}(N,M);P\mid \texttt{in}(N,x);Q&\rightarrow&P\mid Q\{^M/_x\}&\\
\texttt{let}\; x=D\; \texttt{in}\; P\; \texttt{else}\; Q &\rightarrow& P\{^M/_x\}& \\
\multicolumn{4}{r}{if D\Downarrow M}\\
\texttt{let}\; x=D\; \texttt{in}\; P\; \texttt{else}\; Q &\rightarrow& Q& \\
\multicolumn{4}{r}{\textnormal{if there is no }M\textnormal{ such that }D\Downarrow M}\\
!P&\rightarrow&P\mid !P&\\
\end{array}
\end{displaymath} 
\begin{displaymath}
\begin{array}{l l } 
\infer[]{P\mid R\rightarrow Q\mid R}{P\rightarrow Q}&\infer[]{\texttt{new}\; a;P\rightarrow \texttt{new}\; a;Q}{P\rightarrow Q}\\
&\\
\multicolumn{2}{c}{\infer[]{P'\rightarrow Q'}{P'\equiv P,P\rightarrow Q, Q\equiv Q'}}\\
\end{array}
\end{displaymath} 
We write $\rightarrow^*$ for the reflexive and transitive closure of $\rightarrow$. 
\subsection{Observational Equivalence}
A context is a process with a hole, $C[\_]$. An \textit{evaluation context} is a context whose hole is not under a replication, a conditional, an input or an output. An evaluation context represents an attacker, i.e. any process which may follow or not the protocol rules and interact with our ``well behaving'' processes. We say that a context closes $A$ when $C[P]$ is closed, where $C[P]$ is the process obtained by filling $C$'s hole with $P$. We write $P\downarrow _M$ if $P$ can send a message on the channel $M$, that is, if $P\rightarrow^* C[out(M, N);P']$ for some term $N$ and some evaluation context $C[\_]$ that does not bind the free names in $M$
\begin{defn}
\textit{Observational equivalence} ($\approx$) is the largest symmetric relation $\mathcal{R}$ between closed processes with the same domain such that $P\;\mathcal{R}\;Q$ implies:
\begin{enumerate}
 \item  if $P\downarrow _M$ then $Q\downarrow _M$
 \item if $P\rightarrow^*P'$ then $Q\rightarrow^*Q'$ and $P'\;\mathcal{R}\;Q'$ for some $Q'$
 \item $C[P]\;\mathcal{R}\;C[Q]$ for all closing evaluation context $C[\_]$
\end{enumerate}
\end{defn}
Intuitively, observational equivalence captures the idea that two processes are indistinguishable if their observable behaviour is the same. The observable behaviour of a process is what an attacker (evaluation context) can observe and deduce using the equivalence relations on terms when the process interacts with the environment i.e. outputs on free channels. 

The \texttt{ProVerif} process language can represent pairs of processes having the same structure and differing only by some terms and term evaluations, such a pair of processes is called biprocess. Biprocesses are represented by extending the language grammar with the $\texttt{choice}[M,M']$ term and the $\texttt{choice}[D,D']$ term evaluation. A biprocess $P$ defines two processes: $fst(P)$ which is obtained by replacing all occurrences of $\texttt{choice}[M,M']$ with $M$ and $\texttt{choice}[D,D']$ with $D$ in $P$, and, $snd(P)$ obtained by replacing $\texttt{choice}[M,M']$ with $M'$ and $\texttt{choice}[D,D']$ with $D'$ in $P$. 

The \texttt{ProVerif} tool can prove the observational equivalence of biprocesses, which is defined as follows:
\begin{defn} Let $P$ be a closed biprocess. We say that $P$ satisfies observational equivalence when $fst(P) \approx snd(P)$.\end{defn}
\section{Fixing the UMTS AKA Protocol}
In this section, we present the verification of the unlinkability and anonymity properties of the UMTS AKA protocol when sending the same error message for both MAC verification and synchronization failures. We use the formalization of privacy related properties as given by Arapinis et al. in \cite{Arapinis10:AUAUAPC}, namely strong unlinkability and strong anonymity. The definitions we present here are tailored to our case study, we refer the reader to  \cite{Arapinis10:AUAUAPC} for the general ones.

Though the theory allows to write a set of reduction rules to model the \texttt{xor} function, the \texttt{ProVerif} tool cannot deal with its algebraic properties. Hence, in the models used for the verification of unlinkability and anonymity the \texttt{xor} is replaced in the SN and MS models presented in example \ref{exproc}, in the following way:
\begin{itemize}
  \item The \texttt{xor} in the calculation of \texttt{autn} by the SN process at line 5 is substituted by a pair:\\
  \verb|let autn = ((sqn, ak), mac) in|
  \item The \texttt{xor} at line 6 of the MS process is replaced by a pattern matching on the received pair:\\
  \verb|let (xsqn, xak) = y in|
\end{itemize}
The fixed version of the SN and MS process, sending the same error message for both MAC verification and synchronization failures and without the use of the \texttt{xor} function, is shown in figure \ref{snms}.
\begin{figure}
 \begin{verbatim}
let SN = 
    new rand;
    let mac = f1(k, (rand, sqn)) in
    let ak = f5(k, rand) in
    let autn = ((sqn, ak), mac) in
        out(c, (rand, autn));
        in(c, xres).

let MS =   
       in(c, x);
       let (xrand, xautn) = x in
       let (y, xmac) = xautn in
       let ak = f5(k, xrand) in
       let y = (xsqn, xak) in
       let mac = f1(k, (xrand, xsqn)) in
         if xmac = mac then
             if xsqn = sqn then
                let res = f2(k, xrand) in
                   out(c, res)
             else 
                out(c, Fail)
         else
             out(c, Fail).
 \end{verbatim}
  \caption{SN and MS process, sending the same error message for both MAC verification and synchronization failures and without the use of the \texttt{xor} function}
\label{snms}
\end{figure}
\subsection{Strong Unlinkability}
Informally, the strong unlinkability property holds when a system in which agents access a service multiple times looks the same as a system in which agents access the service at most once.  We use this definition since it captures well the concept of user untraceability stated in the 3GPP standard.
Let $SN$ and $MS$ be a serving network process and a mobile station process as defined in example \ref{exproc} and consider the system $S_{UNLINK}$ defined as follows:
\begin{center}            
$S_{UNLINK}=!\texttt{new}\; k;\texttt{new}\; imsi;\texttt{new}\; sqn;(SN\mid\; MS)$
\end{center}
where $k$ is the long term shared key. Each instance of $MS$ in $S_{UNLINK}$ represents a different mobile station, with key $k$ and identity $imsi$ which can execute the protocol at most once. The system $S_{UNLINK}$ is an ideal system with respect to the unlinkability of a mobile station, which by definition cannot be linked because it executes only once.\\
Let $S$ be a system where mobile stations can execute more than once, as defined in example \ref{exproc}.
Following the definition given in \cite{Arapinis10:AUAUAPC}, the unlikability property is satisfied if the system $S$ is observationally equivalent to a system $S_{UNLINK}$:
\begin{center}        
$S$ $\approx$ $S_{UNLINK}$
\end{center}
\subsubsection{ProVerif Verification}
To make the defined observational equivalence suitable for the verification with \texttt{ProVerif}, we defined the two systems $S$ and $S_{UNLINK}$ as the following biprocess $\overline{S}$:
\begin{displaymath}
\begin{array}{r c l}
\overline{S}&=&!\;\texttt{new}\; sk1;\texttt{new}\; imsi1;\\
&&!\;\texttt{new}\; sk2;\texttt{new}\; imsi2;\texttt{new}\; sqn;\\
&&\texttt{let } (k, imsi) = \texttt{choice}[(sk1, imsi1), (sk2, imsi2)] \texttt{ in} \\
&&(SN\mid\; MS)
\end{array}
\end{displaymath}
where $sk1,sk2$ are long term keys and $imsi1, imsi2$ are long term identities. We have that $fst(\overline{P})$ represents a system where a mobile station (with key $sk1$) may execute the protocol many times, while $snd(\overline{P})$ represents a system where mobile stations execute the protocol at most once (the key $sk2$ is always different). This makes the definition of strong unlikability suitable for the automatic verification using \texttt{ProVerif}, because it reduces the problem of testing strong unlinkability to the observational equivalence of a biprocess ($fst(\overline{P})$ and $snd(\overline{P})$ differ only in the choice of the identity and related key, i.e. $\texttt{choice}[(sk1, imsi1), (sk2, imsi2)]$).

The verification of the simple fix of the UMTS AKA protocol with the \texttt{ProVerif} tool shows that the unlinkability property holds. The verification of the original protocol results, as expected, in the breach of the unlinkability property. The \texttt{ProVerif} code used to verify the unlinkability property of the original protocol and of the fixed version is shown respectively in figure \ref{provUn} and figure \ref{simplefix}.
\subsection{Strong Anonymity}
Informally, strong anonymity requires a system where the user $imsi_w$ with publicly known identity executes the role $R_i$ to be undistinguishable from a system where the user $imsi_w$ is not present at all, which is the ideal system from $imsi_w$'s anonymity preservation point of view. This notion captures well the definition of user identity confidentiality stated in the 3GPP standard.

Let $SN$ and $MS$ be a serving network and a mobile station process as defined in example \ref{exproc} and let $MS_W$ be mobile station processes such that. $MS_W$ has known, public identity $imsi_w$. $MS_W$ is defined as follows:
\begin{displaymath}
 \begin{array}{r c l}
  MS_W&=&!MS\{^{imsi_w}/_{imsi}\}\\
 \end{array}
\end{displaymath}
where $imsi_w$ is the publicly known identity of $MS_W$. The strong anonymity holds when the following observational equivalence is satisfied:
\begin{center}
$!MS\mid\; !SN$ $\approx$ $!MS\mid\; MS_W\mid\;!SN$
\end{center}
\subsubsection{ProVerif Verification}
The mobile stations' permanent identities are modelled as secret names, to verify the strong anonymity property we give the adversary the knowledge of one identity, modelled as a free (public) name. We define a system $S_W$ where the mobile station with publicly known identity $imsi\_w2$ executes the protocol as follows: 
\begin{displaymath}
 \begin{array}{l c l}
  \texttt{free}&\multicolumn{2}{l}{imsi\_w2.}\\
  S_W&=&\texttt{new}\;sk\_w2;\;\\
		&&!\texttt{new}\;sk\_ms;\texttt{new}\;imsi\_ms;\texttt{new}\;sqn\_ms;\texttt{new}\;sqn\_w;\\
		& &\texttt{let } (imsi,k) = (imsi\_ms, sk\_ms) \texttt{ in} \\
		&&\texttt{let } sqn = sqn\_ms\texttt{ in }(SN \mid MS)\\
		&&|\\
		&&\texttt{let }(imsi,k) = (imsi\_w2, sk\_w2)\texttt{ in}\\
		&&\texttt{let }sqn = sqn\_w\texttt{ in }(SN\mid MS)
 \end{array}
\end{displaymath}
where $sk2$ and $sk\_ms$ are permanent shared keys, $imsi\_w2, imsi\_ms$ are permanent mobile stations identities. The given model represent a system where a mobile station with public identity $imsi\_w2$ can run the protocol in parallel with unboundedly many mobile stations having secret identities. 
The anonymity properties as tested by \texttt{ProVerif} is described by the following biprocess:
\begin{displaymath}
 \begin{array}{l c l}
  \multicolumn{2}{l}{\texttt{free}}&imsi\_w2.\\
  \overline{S}&=&\texttt{new}\;sk\_w1;\;\texttt{new}\;imsi\_w1;\;\texttt{new}\;sk\_w2;\;\\
		&&!\texttt{new}\;sk\_ms;\texttt{new}\;imsi\_ms;\texttt{new}\;sqn\_ms;\texttt{new}\;sqn\_w;\\
		& &\texttt{let } (imsi,k) = (imsi\_ms, sk\_ms) \texttt{ in} \\
		&&\texttt{let } sqn = sqn\_ms\texttt{ in }(SN \mid MS)\\
		&&|\\
		&&\texttt{let }(imsi,k) =\\
		&&\hspace{3mm}\texttt{choice}[(imsi\_w1, sk\_w1),(imsi\_w2, sk\_w2)]\texttt{ in}\\
		&&\texttt{let }sqn = sqn\_w\texttt{ in }(SN\mid MS)
 \end{array}
\end{displaymath}
where $fst(\overline{S})$ is a system where only mobile stations with secret identities execute the protocol and $snd(\overline{S})= S_W$ is a system where the mobile station with publicly known identity $imsi\_w2$ may run the protocol.

The relevant parts of the \texttt{ProVerif} code used to verify the anonymity property are shown in figure \ref{provAnon}.

\texttt{ProVerif} proved that the anonymity property is achieved by both the fixed version and the original version of the UMTS AKA protocol.
\section{Discussion}
The \texttt{xor} function is used in the UMTS AKA protocol to combine the sequence number $SQN$ with the $AK$ parameter. Instead of modelling a simplified version of the \texttt{xor} function, we chose to send the SQN and AK parameter in clear, i.e. as components of a tuple. We believe this abstraction is sound, in this case, because it gives more discretional power to the attacker. However, a formal proof of the soundness of this abstraction is not given in this paper. 

Besides the xor abstraction, the verification of the fixed version of the AKA protocol shows that our simple fix thwarts the information leakage problem which causes the linkability of UMTS subscribers. 

\begin{figure}
\begin{tikzpicture}[node distance=4cm, auto,>=latex', thick]
	
    \path[use as bounding box] (0.8,3) rectangle (16,-3.5);
    \draw[elec] (5.5,2.5) node[format](ms){\small $K$, $SQN_{MS}$, $PK_{HN}$}-- (5.5,-3.5);
    \draw[elec] (11,2.5) node[format](sn){\small $K$, $SQN_{HN}$, $SK_{HN}$}-- (11,-3.5);
	\node [above=3mm] at (sn){SN/HN};
    \node [above=3mm] at (ms){MS};
	\node at (11.1,1.3)(startreq){};
    \node at (5.4,1.3)(endreq){};
    \path[elec,->](startreq) edge node[above] {\footnotesize AUTH\_REQ$(RAND,AUTN)$} (endreq);
	\draw[elec] (5.5,-0.8) node [format, fill=white](sn_pre){
			\parbox{4.2cm}{
				{\footnotesize
 				if $MAC\neq XMAC$ then \\
				$RES\leftarrow \{MAC\_FAILURE\}_{PK_{HN}}$\\
				else \\
				if $SQN_{HN}<=SQN_{MS}$ then\\
				$RES\leftarrow\{SYNCH\_FAILURE\}_{PK_{HN}}$\\
				else $RES\leftarrow f2_K(RAND)$
				}
			}
	};
    \node at (5.4,-3)(startresp){};
    \node at (11.1,-3)(endresp){};
    \path[elecred,->](startresp) edge node[above] {\footnotesize AUTH\_RES$(RES)$} (endresp);
\end{tikzpicture}
\caption{AKA protocol with public key encryption of error messages}
\label{fixpublickey}
\end{figure}
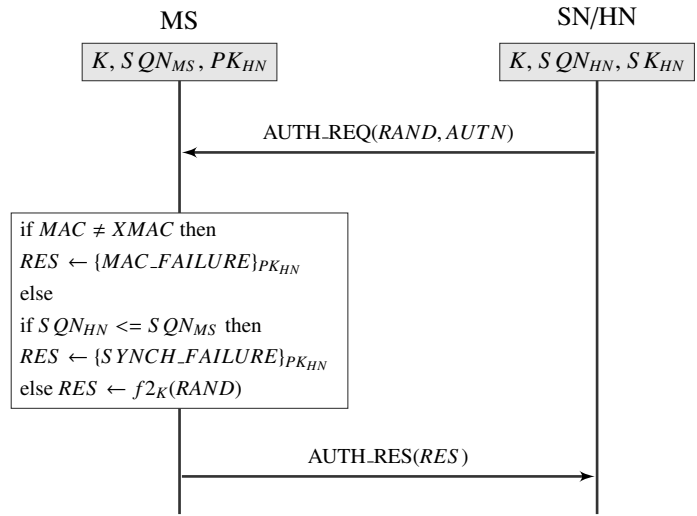

The proposed simple solution does not diversify the type of errors, hence it would invalidate error recovery features of the protocol. Instead, we suggest the adoption of a public key infrastructure where each home network has a public key known by the mobile stations (it could be stored in the USIM). When an error occurs, during the authentication procedure, the mobile station can encrypt the error message using the HN public key and then send the error message on the radio channel (see figure \ref{fixpublickey}). Notice that the introduction of random padding values will be required to obtain a non-deterministic encryption of the error messages and avoid known cypher text attacks being performed by the attacker.
Furthermore, meaningful error messages allowing re-synchronization and sender authentication should be used in a real implementation.
\section{Future Work}
The work presented in this document is still in progress. We discuss in this session the issues currently under examination.
In modelling the AKA protocol, we made some abstractions to ease the modelling process and to overcome \texttt{ProVerif}'s limitations. In particular, we send the sequence numbers in clear, instead of XOR-ing them with the AK parameter. Moreover, in our model the HN and MS view of sequence numbers is perfectly synchronized.

Intuitively, the XOR abstraction is sound, since we give the attacker more information, with respect to the original protocol, hence we leave more room for attacks. With this abstraction, the unlinkability of the AKA can be proved when our simple fix is applied to it. The perfect synchronization of sequence numbers, intuitively, does not affect the unlinkability property, in fact in both the original AKA protocol and the fixed one, they are sent in clear. Moreover, the linkability attack relies on the possibility of replaying old authentication requests, i.e. old sequence numbers, hence the currently valid sequence numbers are irrelevant from the attacker's point of view. We are currently working on the soundness of our abstractions. 
\texttt{ProVerif} cannot prove the unlinkability of the public key based fix of the AKA protocol a manual proof will be given instead.

While modelling the AKA protocol and studying its privacy properties, we started questioning the feasibility of the linkability attack in real settings. Hence we investigated the available open source projects related to mobile telephony technologies and we are currently working on the actual implementation of the proposed linkability attack.
\section{Conclusions}
We showed that despite the use of temporary identities UMTS subscribers can be traced thanks to the information leaked by the authentication and key agreement protocol. We proposed a public-key-based countermeasure to ensure both unlinkability and anonymity of UMTS subscribers while keeping the UMTS security architecture mostly unchanged. Furthermore, we successfully use the definition given in \cite{Arapinis10:AUAUAPC} to automatically verify privacy related properties. Though our approach is not applicable to more general settings, it can be used to automatically verify unlinkability and anonymity of protocols which do not use the user identity during the initialization phase.
\bibliographystyle{unsrt}
\bibliography{links,FormalTheory,GSMUMTSprivacy,UMTSprivacy,UMTS}
\section{Appendix}
In this section, we present the code used for the verification with \texttt{ProVerif} an clarify some technical details. In the rest of this section we will use the notation:
\begin{verbatim}
(* section heading *)
...
\end{verbatim}
to avoid code repetition and ease the reading.
 
In the following figure, we introduce the functions and reductions we used to model the cryptographic primitives involved in the UMTS authentication and key agreement protocol. 
\begin{figure}[tp]
\begin{verbatim}
(* public communication channel *)
free c.

(* constant values *)
fun macFail/0. fun synchFail/0. fun Fail/0.

(* UMTS AKA protocol specific mac and key generation functions *)
fun f1/2. fun f2/2. fun f3/2. fun f4/2. fun f5/2.

(* generic symmetric encryption function *)
fun senc/2.
reduc sdec(k, senc(k, m)) = m.
\end{verbatim}
\caption{The public channel, the cryptographic primitives and the constant values}
\label{fun}
\end{figure}
The function \texttt{senc} model symmetric encryption, while \texttt{f1,f2, f3, f4} and \texttt{f5} are protocol specific one way cryptographic functions. Moreover, in this part of the code we declare a public channel \texttt{c} and three constant values \texttt{macFail}, \texttt{synchFail} and \texttt{Fail}. The public channel and function declaration is reported in figure \ref{fun}.
\begin{figure}[tp]
  \begin{verbatim}
(* Serving Network process *)
let SN = new rand;
    let mac = f1(xk, (rand, xsqn)) in
    let ak = f5(xk, rand) in
    let autn = (xsqn, ak, mac) in
        out(c, (rand, autn));
        in(c, xres).
  \end{verbatim}
\caption{Serving Network Process}
\label{sn}
\end{figure}
The serving network process did not present any technical challenge in the modelling process and is the same for all the verified properties, hence is briefly reported in figure \ref{sn}, without further remarks.

The code in figure \ref{provUn} was used to verify the unlinkability of the UMTS AKA protocol. Though, the evaluation of an \texttt{if} statement condition is part of the internal behaviour of a process, \texttt{Proverif} considers processes, which execute different branches of an \texttt{if} statement, not observationally equivalent. Hence it might exhibit false attacks. In fact, the \texttt{ProVerif} tool is proved to be sound but not complete. This means that if \texttt{ProVerif} does not find an attack, this is a proof that there are no attacks, and the security property holds in the verified model. Though, the contrary it is not true, i.e. if \texttt{ProVerif} does find an attack, it is not a proof that the property does not hold, because the attack may or may not match a real attack, and the property could still be proved to be satisfied using, for example, manual techniques. To avoid this problem, when verifying the unlikability property, we introduced a function \texttt{err} and a destructor \texttt{geterr}, which allow to check the validity of MAC and sequence number, and at the same time retrieve the error message. The use of the \texttt{err} function avoids the introduction of extra \texttt{if} statements that would lead to false attacks.
\begin{figure}[tp]
\begin{verbatim}
(* public channel and cryptographic primitives *)
...

fun err/4.
reduc 
   geterr( err(x, z, y, y)) = macFail;
   geterr( err(x, x, y, z)) = synchFail.

(* Serving Network process *)
...

(* Mobile Station process *)
let MS =  in(c, x);
       let (xrand, xautn) = x in
       let (xsqn, xak, xmac) = xautn in
       let mac = f1(k, (xrand, xsqn)) in
          if (xmac, xsqn) = (mac, sqn) then
             let res = f2(k, xrand) in
                out(c, res)
          else 
             let errmsg = geterr(err(mac, xmac, sqn, xsqn)) in 
                out(c, errmsg)
     ).

process  !new sk1; new imsi1;
         !new sk2; new imsi2; new sqn;
         let (k, imsi) = choice[(sk1, imsi1), (sk2, imsi2)] in
         (SN | MS)
\end{verbatim}
\caption{\footnotesize AKA protocol Unlinkability}
\label{provUn}
\end{figure}
As explained in section \ref{linkatt}, the UMTS AKA protocol does not satisfy the linkability property and in fact \texttt{ProVerif} cannot prove the observational equivalence and finds that an attacker can distinguish the two processes because different error messages can be sent as answer for the same authentication request.

The code in figure \ref{simplefix} shows the AKA protocol fixed so that the same error message is sent, independently from the kind of failure occurred. This is obtained by simply changing the definition of the err function. This solution is proved to satisfy unlinkability.
\begin{figure}[tp]
\begin{verbatim}
(* public channel and cryptographic primitives *)
...

fun err/4.
reduc 
   geterr( err(x, z, y, y)) = Fail;


(* Serving Network process *)
...

(* Mobile Station process *)
...

process  !new sk1; new imsi1;
         !new sk2; new imsi2; new sqn;
         let (k, imsi) = choice[(sk1, imsi1), (sk2, imsi2)] in
         (SN | MS)
\end{verbatim}
\caption{\footnotesize AKA protocol Unlinkability: simple fix}
\label{simplefix}
\end{figure}
In figure \ref{provAnon}, we show the code we used to verify the AKA anonymity property. In this case, the if statements were not source of false attack, hence we did not use the \texttt{err} function. As expected the anonymity property holds because the private identities of the mobile stations are never revealed (i.e. sent on the public channel).
\begin{figure}[tp]
\begin{verbatim}
(* public channel and cryptographic primitives *)
...

(* Serving Network process *)
...

(* public identity *)
free imsi_w2.

(* Mobile Station processes *)
let MS =
    in(c, x);
    let (xrand, xautn) = x in
    let (xsqn, xak, xmac) = xautn in
    let mac = f1(k, (xrand, xsqn)) in   
    if (xmac, xsqn) = (mac, sqn) then 
        if xsqn = sqn then 
            let res = f2(k, xrand) in
                out(c, res)
        else out(c, synchFail)
    else out(c, macFail).

process new sk_w1;new sk_w2;new imsi_w1;
        !new sk_ms;new imsi_ms;
        !new sqn_w;new sqn_ms;
        let (imsi,k) = (imsi_ms, sk_ms) in 
            let sqn = sqn_ms in (SN |MS)
        |
        let (imsi,k) = 
            choice[(imsi_w1, sk_w1),(imsi_w2, sk_w2)] in
            let sqn = sqn_w in (SN|MS)
\end{verbatim}
\caption{\footnotesize AKA protocol Anonymity}
\label{provAnon}
\end{figure}
\end{document}